# A SOFTWARE REUSE APPROACH AND ITS EFFECT ON SOFTWARE QUALITY, AN EMPIRICAL STUDY FOR THE SOFTWARE INDUSTRY

**Ahmed Mateen**[*]

**Samina Kausar**[*]

**Ahsan Raza Sattar**[*]

**Abstract**

Software reusability has become much interesting because of increased quality and reduce cost. A good process of software reuse leads to enhance the reliability, productivity, quality and the reduction of time and cost. Current reuse techniques focuses on the reuse of software artifact which grounded on anticipated functionality whereas, the non-functional (quality) aspect are also important. So, Software reusability used here to expand quality and productivity of software. It improves overall quality of software in minimum energy and time. Main objective of this study was to present a reuse approach that discovered that how software reuse improves the quality in Software Industry. The V&V technique used for this purpose which is part of software quality management process, it checks the quality and correctness during the software life cycle. A survey study conducted as QUESTIONAIR to find the impact of reuse approach on quality attributes which are requirement specification and design specification. Other quality enhancement techniques like ad hoc, CBSE, MBSE, Product line, COTS reuse checked on existing software industry. Results analyzed with the help of MATLAB tool as it provides effective data management, wide range of options, better output organization, to check weather quality enhancement technique is affected due to reusability and how quality will improve.

[*] Department of Computer Science, University of Agriculture, Faisalabad, Punjab, Pakistan

# 1. Introduction

Software reuse is a way toward making software frameworks by available elements as opposed to building software frameworks without any work from initial stage. It is currently acknowledged that for accomplishing enhanced software, more quickly and at lower cost, there is a requirement to implement a design process which grounded on orderly reuse of software. Forming reusable software assets requires a settled organization whose developers and architects can distinguish key sources of variability in their application domain [1]. Software components available with the collections provided with that software. These software parts are known as components. These components can be used at any time when needed [2].From many years, research scholars and software developers define the relationship in the development of new software and software reuse is as similar as the relation of development of new products with the idea of earlier products [3].

As compared to non-reused a reused component produces low defect rate. The defect solidity is calculated by total defects divided by total number of line of code. However with high level of rigorousness, more defects encountered with reused components than total division, but after delivery these defects resulted very low.

It is an inspiration for researchers to reuse the software applications for collaborating so as to upgrade quality and minimize software cost. Conversely, merely a partial percentage of the prototypes of reuse presents operative indication in real according to their suitability and operative reuse is not easy to usability and usefulness [4].If a work is already done than there is no need to spent cost on that again except its weaknesses and new improvements. Quality attribute must be correct, ranked, complete, verifiable, consistent, modifiable and usable [5]. Various products of software are industrialized by using ongoing types or arrangements for minimizing time of delivery of software product in order to increase the output and worth and also for reducing struggles [6].

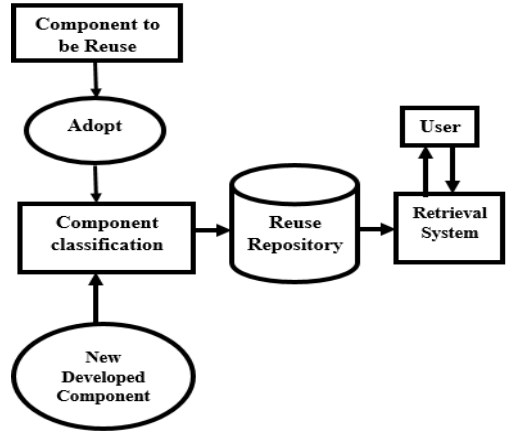

Figure: 1.1.*Component reuse*

1.1. Software quality

The qualityof a software item is the way the framework under operation satisfies the necessities and prerequisites as characterized (certainly or unequivocally) by its stake holders." And "Quality assurance is the set of exercises completed to guarantee that the framework has adequate quality." To ensure quality it is checked that the product is working what for it is designed i.e. validation and verification. But the software quality can't be viewed as separated, yet rather it is set of joined exercises resulting in desired output that take place in various phases of software life cycle [7].

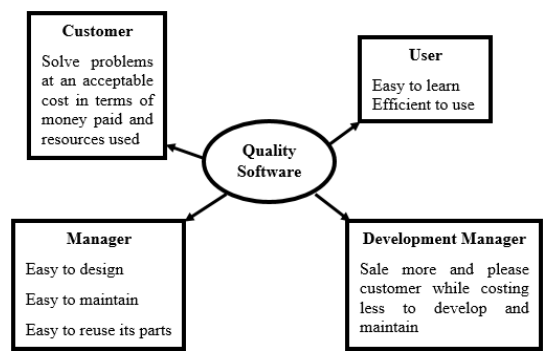

Figure: 1.2.*Software Quality*

## 2. Literature Review

A high quality software reuse process contributed to improved productivity, quality and dependability, in addition to aiding the acquisition professional in better management of the schedule, cost and performance of a program or project. An initial investment is necessary which can be relatively paid for itself with the passage of time. Development of a reuse program and the reuse process employed in that program can benefit both to reduce risking factors and new system expansions. [8]

Four different levels of analysis influenced effect on the reuse success. That were: strong, weak, none and no data, that component has some impact on programming reuse. Survey study analysis showed that for the software companies it is a key feature to reuse the software concerned for the betterment of quality and budget and should have low risk [9]. High complexity caused difficulties. The reuse design principles must be clear, precise interface and understandable it will increase the ease of reuse significantly [10]. General shortage of knowledge appeared about the relationship among quality assurance and software systems. So here conveyed a preliminary description of the specification in this region [11].

Few computer software reuse tasks involve reusing of code. These kinds are as, (1) significantly minimize some time that programmers demand to accomplish practical reuse tasks, (2) increase the chances that programmers can efficiently total practical reuse tasks, (3) minimize some time essential by programmers to name infeasible reuse tasks, in addition to (4) enhance developers' good sense in their capability to control danger such tasks[12]. For the quick development, minimized cost and decreasing risk factor and best use of developers, software reuse contains high benefits. At the early stages of developing software reusing results in increasing benefits as it allows the previous objects to be use in the successive reuse of later stage objects [13].

In different viewpoint when a feature reused, the weaver should be able to apply the aspect to target program according to dissimilar settings. The configurability of aspect mainly defined the point cuts, assignment of information and coordinating the relationship between different features [14]. Benefits resulted by reuse are as under: firstly, blended problems disabled, secondly much flexibility achieved about what and when specifications should be reused and finally the third one is that the developer did not have any need of deep knowledge about full path of reusing objects.[15].

Quality is serious to the influences of selection, maintenance, and service quality evaluation to a great range. The quality evaluation technique defined all aspects by founding an attribute model and categorizes domain software quality by forming level model [16].

## 3. Objective

Main goal of this exploration is to present a reuse approach that will discover how software reuse improves the quality in Software Industry. The V&V technique used for this purpose. A survey study conducted like QUESTIONAIR to find the impact of reuse approach on quality attributes which will be requirement specification and design specification. Other quality enhancement techniques like ad hoc, CBSE, MBSE, Product line, COTS reuse will be checked on existing study. The result shows that reuse enhance the quality of software and reduces the cost.

## 4. Methodology

A survey conducted to check the consequences of new approach developed by mixing up the previous different approaches as ad hoc, component reuse and model based reuse etc. The effect of this reuse approach checked through a survey. Quality attributes like requirements and design specification checked. Verification and Validation technique apply that checks the quality and correctness during the software life cycle. The survey is done to ensure software Quality Attributes, while using standards for safety critical software systems and selected Verification and Validation techniques.

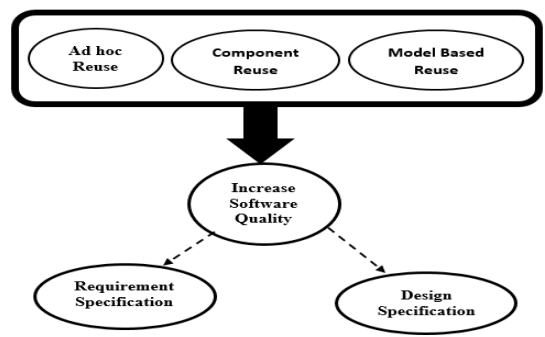

*Methodology Overview*

To develop the proposed approach following steps performed.

- Review different old approaches
- Understanding what is software reuse
- Understanding the relationship between quality SR
- Select the quality attributes to analyzing
- Purpose a new approach
- Define the steps of new approach
- Other techniques compared
- Survey conducted
- Risk factor analyzed
- Result analyzed

Software quality achievement is the basic purpose of this research. According to customer quality is to fulfill his requirement at an acceptable level of cost. According to user quality is easy to use and work with software. According to developer quality is easy to design, maintain and reuse software's parts and according to developer quality is low development cost as well as high customer satisfaction. To fulfill these demands our approach will use a questionnaire that filled by different software related persons. Here represented some quality parameters.

### 4.1. Requirement specification

The proposed model focused on the defects in the requirements specification phase and further correlation between the weakness in requirement specification and its roots to assure the quality of the SRS document. The technique helps in training the SRS authors, and supports creating more exact and correct fault prevention and fault detection techniques [17]. The necessities may affected from deficiency and the presence of various and irregular words exists in SRS due to the authoritative, multicultural and singular prerequisites. The rate of SRS in light of class properties arranges the protest techniques as rehashed explanations [18].

### 4.2 Design specification

The design stage executes a critical part in connecting the necessities stage and the usage stage, amid software development phases and it is a method of converting necessities of software interested in software societies. The methodology self-possessed of a software strategy description practice and a method of software design consideration. Reliable with the said design topographies, a design examination method has been planned. Recognized properties, such as correctness, consistency, extensiveness and reusability could be certified in design level [19].

Verification and validation are two important terms that are used in the industry of testing or quality insurance. They both seem to the same thing. However, they both are little different when it comes to the world of software product. Software or any related product needs to go through the complete cycle of development and testing before being launched in the market. Verification evaluates all the necessary items related to the product being developed. It is important to mention that during verification, one is not testing the actual product.

This is new study for all organization frameworks and programming designers to decide different encounters they had with various sorts of reuse. This data can be utilized to examine and relate finest applies for the reuse. It can likewise be utilized as information toward incorporate as a part of recommendations, for research papers and experts proposals. Following motivations exists behind this review:

- A methodology is the decision of approach or mix of methodologies the program marks to utilize reuse

- A method is the unique from the improvement strategies that takes into account reuse

- Component-constructed reuse is situated in light of effectively created parts or intended for reuse on a segment premise.

- Model-based reuse will be reuse that depends on reusing models made from different projects or segments.

- Product line reuse will be reuse in light of an institutionalized yet tailorable product offering.

- Ad Hoc reuse that the specialist knows about, that happens to meet a prerequisite yet was not intended for reuse.

### 2.2.1  Respondent information

The reason for these queries is to associate reuse involvement with the kind of architect, programming, frameworks), the organization, and the experience level of the architect

### 4.2.2 Reuse information

This arrangement of queries distinguishes the sort of reuse technique utilized, whether achievement is enhanced with being a piece of the choice, and whether the software is sufficiently far along to gauge considers that happen late the program. So improvement procedures looked at and ancient rarities utilized on programming frameworks.

### 4.2.3 Reuse effectiveness information

This arrangement of queries will associate the adequacy of the system against the methodology by distinguishing and counting the adjustment in results credited to reuse.

Participant included in the survey were: software administrators, software investigators, framework specialists and software engineers. By gathering responses from experts with various foundations, to investigate alternate points of view, and how reuse includes specialized and non-specialized perspectives.

### 5. Results

To carry out this research 25 questions were made by analyzing the previous history of software reuse to check the quality attributes as design specification and requirement specification. Different software houses were visited and they answered the survey. The respondents of the survey included as under: software systems engineer, IT managers, software engineers and software developers. The answers collected from the different background having respondents, they provide different views and analysis and in what way reuse includes practical opinions. About 50% defendants (28 contributors) appealed to attain achievement in software schemes owed to reuse of the software which is a great advantage.

89.8% male and 10.2% female respondents encountered from total 50 respondents. Type of engineers were mostly software engineers and 98 % responses about their experience.

The following table shows the respondent information that what type of engineer they were. So it can be checked that 78 % software and 12% system engineer participate to the survey.

### 5.1. Software Reuse effects on quality

The following da ta collected to know the effect of software reusability on the quality of software. As result of survey shows that according to 11.8% engineer reuse has no effect on quality. Whereas, 86.3% answered in yes. Which means software reuse has high impact on the quality of software.

### 5.2. Main focus of developers during reuse

Question was that what the focus of developers is during reusing the software. According to response component base software reuse preferred which was 40%. There is no higher difference between other three parameters. As 38 and 22 percent respectively.

### 5.3. Chi square test on the research question

A statistical test applied to the survey result which is chi square test, in which two variables testes based on the following hypothesis.

H0: Mixed reuse strategies approach has positive effect on software quality.

H1: Mixed reuse strategies approach does not have positive effect on software quality

Table 5.1.Effects of Software Reuse on quality

|       |       | Frequency | Percent | Valid Percent | Cumulative Percent |
|-------|-------|-----------|---------|---------------|--------------------|
| Valid |       | 1         | 2.0     | 2.0           | 2.0                |
|       | No    | 6         | 11.8    | 11.8          | 13.7               |
|       | yes   | 44        | 86.3    | 86.3          | 100.0              |
|       | Total | 51        | 100.0   | 100.0         |                    |

Table 5.2. Developer focus while reuse

| | | Frequency | Percent | Valid percent | Cumulative percent |
|---|---|---|---|---|---|
| Valid | Completesoftware | 19 | 37.3 | 38.0 | 38.0 |
| | Component | 20 | 39.2 | 40.0 | 78.0 |
| | Both | 11 | 21.6 | 22.0 | 100.0 |
| | Total | 50 | 98.0 | 100.0 | |
| Missing | System | 1 | 2.0 | | |
| Total | | 51 | 100.0 | | |

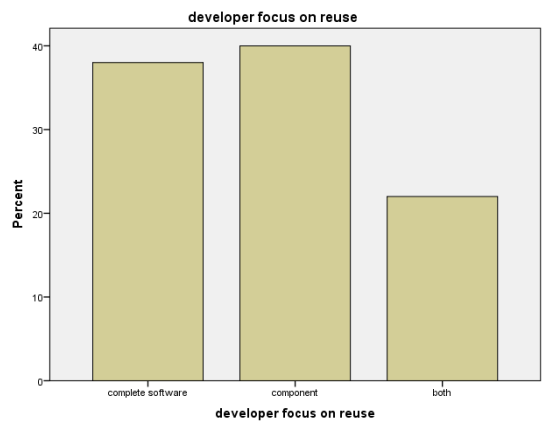

Figure 5.1.*Developer focus on Reuse*.

Table 5.3.Chi square test for quality

**Chi-Square Tests**

| | Value | DF | Asymp. Significance (2-sided) | Exact Significance (2-sided) | Exact Significance (1-sided) |
|---|---|---|---|---|---|
| Pearson Chi-Square | 4.715[a] | 1 | .030 | | |
| Continuity Correction | 2.986 | 1 | .084 | | |
| Likelihood Ratio | 6.865 | 1 | .009 | | |
| Fisher's Exact Test | | | | .069 | .034 |
| N of Valid Cases | 49 | | | | |

Table 5.4.Proposed Approach effect on Quality

|  |  | Reuse mix approaches effect on quality | | Total |
|---|---|---|---|---|
|  |  | no | Yes |  |
| **Reuse improve quality** | No | 0 | 6 | 6 |
|  | Yes | 20 | 23 | 43 |
| **Total** |  | 20 | 29 | 49 |

Reuse improve quality * Reuse mix approaches effect on quality Cross tabulation

a. 2 chambers (50%) have predictable amount less than 5. The least predictable amount is 2.45.

b. Calculated simply for a 2x2 table

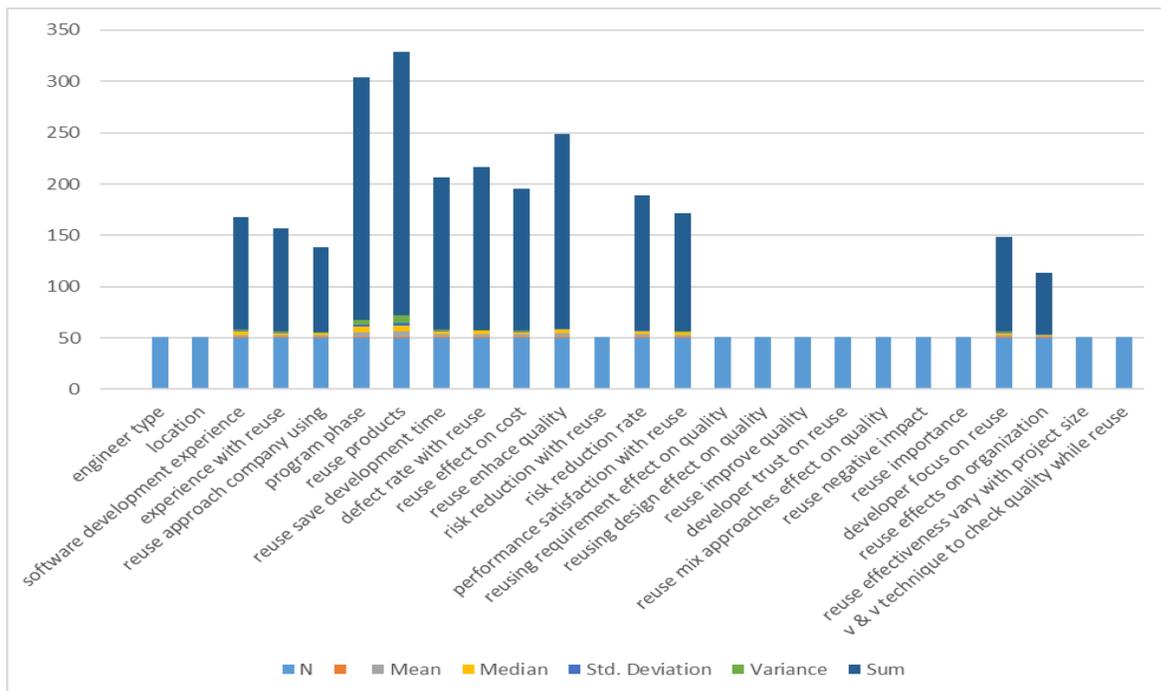

Figure 5.2.*Overall Results*

Figure 5.2 shows the overall complete results of the survey with the help of mean, median. Standard deviation, variance and sum.

5.4. Overall Results of the Survey

So there exist significant relationship between software quality and mixed reuse approach.

## 6. Conclusion

In this paper a survey conducted to check the result of software reusability for the value of the software which as 71% resulted in that point of view that software reuse approaches combine effect better to enhance the quality, efficiency, suitability,maintainability, and reduces the time and cost which is a great benefit. However, it is suggested for future researchers to enhance this research as in this work the sample size is very low and data collected is small. If the size of data is small it will help in understandings more effectively and clearly.

## References

[1]     BakarN. H., Kasirun Z. M. and SallehN., "Feature extraction approaches from natural language requirements for reuse in software product lines: A systematic literature review"*Journal of Systems and Software*, *106(1)*, 132-149, 2015.

[2]     Aggarwal D. and NaveetaM., "Software reuse- A compendium"*International Journal of Research in IT and Management, 2(2)*, 93-100, 2012.

[3]     CusumanoM. A., "Japan Software Factories: A Challenge to U. S. Management"*Oxford University Press, New York,* 230-241, 2006.

[4]     BuccellaA., CechichA., PolM., AriasM., Doldan M. D. S. and MorsanE., "Marine ecology service reuse through taxonomy-oriented SPL development"*Journal ofComputers & Geosciences,73(1),* 108-121, 2014.

[5]     KrogerT. A., Neil J. D. and StephenC. C., "Understanding the characteristics of quality for software engineering processes: A Grounded Theory investigation"*Journal of Information and Software Technology,56(1),* 252-271, 2014.

[6]     ParakashB. V. A., Ashoka D. V. and ManjunathA. V. N., "Application of Data Mining Techniques for Software Reuse Process"*Journal of Procedia Technology*, *4(1),* 384-389, 2012.

[7] JansenS., and van CapelleveenG., "Quality review and approval methods for extensions in software ecosystems"*Software Ecosystems: Analyzing and Managing Business Networks in the Software Industry*, 187, 2013.

[8] GreathouseC. A., "Software reuse in the Naval Open Architecture"*Doctoral dissertation, Naval Postgraduate School 1(1),* 55-150, 2008.

[9] LucredioD., SantosB., AlvaroK., GarciaA., AlmeidaV. C. D., MattosE. S. D., Fortes R. P. and MeiraS. L., "Software reuse: The Brazilian industry scenario"*Journal of Systems and Software*, *81(6),* 996-1013, 2008.

[10] AnguswamyR., and FrakesW. B., "A study of reusability, complexity, and reuse design principles. In Empirical Software Engineering and Measurement (ESEM)" *ACM-IEEE International Symposium, 1(1),* 161-164, 2012.

[11] BabaderA., RenJ., Jones K. O. and WangJ., "A system dynamics approach for enhancing social behaviours regarding the reuse of packaging"*Journal of Expert Systems with Applications*, *46(1),* 417-425, 2016.

[12] HolmesR., and RobertW. J., "Semi-Automating Pragmatic Reuse Tasks" *Laboratory for Software Modification Research University of Calgary Calgary, Alberta, Canada. IEEE Conference,1(1),* 481-482, 2008.

[13] SalamiH. O., and AhmedM. A., "UML artifacts reuse: state of the art"*arXiv preprint arXiv,1(1),* 1402-0157, 2014.

[14] S. Yi, and C. He, "A Comparison of Approaches Toward Reusable Aspects" In *International Conference on Computer Science and Intelligent Communication,1(1),* 1-4, 2015.

[15] DelgadoA.A., Troyano J. A. E. and EstipaR., "Reusing UI elements with Model-Based User Interface Development"*Journal of Human Computer Studies, 86(1),* 48-62, 2015.

[16] Bao, T.and ShufenL., "Quality evaluation and analysis for domain software: Application to management information system of power plant"*Journal of Information and Software Technology,78(1),* 53-65, 2016.

[17] AlshazlyA. A., Ahmed E. M. and MohamedA. S., "Detecting defects in software requirements specification"*Journal of Alexendria Engineering, 53(1),* 513-527, 2014.

[18] ShivaramA. M., and ShivanandH. M., "An Ameliorated Methodology for the Abstraction of Object Oriented Features from Software Requirements Specification"*Journal of Procedia Computer Science,62(1),* 274-281, 2015.

[19]     KooS. R., and SeongP. H., "Software design specification and analysis technique (SDSAT) for the development of safety-critical systems based on a programmable logic controller (PLC)"*Journal of Reliability Engineering and System Safety,91(6),* 648-664, 2006.